# Consistent Quality Oriented Rate Control in HEVC via Balancing Intra and Inter Frame Coding

Wei Gao, Qiuping Jiang, Ronggang Wang, Siwei Ma, Ge Li, and Sam Kwong

*Abstract*—Consistent quality oriented rate control in video coding has attracted much more attention. However, the existing efforts only focus on decreasing variations between every two adjacent frames, but neglect coding trade-off problem between intra and inter frames. In this paper, we deal with it from a new perspective, where intra frame quantization parameter (IQP) and rate control are optimized for balanced coding. First, due to the importance of intra frames, a new framework is proposed for consistent quality oriented IQP prediction, and then we remove unqualified IQP candidates using the proposed penalty term. Second, we extensively evaluate possible features, and select target bits per pixel for all remaining frames, average and standard variance of frame QPs, where equivalent acquisition methods for QP features are given. Third, predicted IQPs are clipped effectively according to bandwidth and previous information for better bit rate accuracy. Compared with High Efficiency Video Coding (HEVC) reference baseline, experiments demonstrate that our method reduces quality fluctuation greatly by 37.2% on frame-level standard variance of peak-signal-noise-ratio (PSNR) and 45.1% on that of structural similarity (SSIM). Moreover, it also can have satisfactory results on Rate-Distortion (R-D) performance, bit accuracy and buffer control.

## I. INTRODUCTION

THE significantly increased amount of video data should be compressed effectively for better visual communication, processing and experience. During the past few years, low complexity [1]-[3], [33] and rate control (RC) [4] algorithms have been intensively developed for fast and high performance video encoders. Undoubtedly, High Efficiency Video Coding (HEVC) [5] is one of the most popular standards for video compression, and its developments and applications have been remarkably improved. Nevertheless, traditional video coding efforts usually focus on better rate-distortion (R-D) gains, rather than frame quality consistency that can also affect visual experience significantly. Also, it is recognized that both spatial and temporal artifacts need to be evaluated [6] in video quality assessment. Thus, quality consistency (QC) is indispensable for comfortable visual experience, and the optimized RC should provide visually pleasant videos with smoother quality.

### A. Quality Consistency Optimization

For quality consistency, different optimization approaches have been reported [7]-[16]. Frame-level RC [18] plays an important role in reducing frame quality fluctuations, where quantization parameters (QPs) are flexibly configured for frames, while block-level adjusting has a negligible impact [17]. Methods for quality consistency can be divided into following categories. The first category is the major approach which uses previous distortion as reference to guide current distortion. For H.264/SVC, Seo et al. [11] defined target distortion of spatial enhancement layer according to base layer, and then derived QP to achieve smooth quality. Xie et al. [12] utilized a nonlinear model to prevent parameter inaccuracy in achieving QPs. Huang et al. [13] proposed to jointly optimize multiple goals, including coding fidelity, quality consistency and bit rate accuracy for inter frames. For HEVC, Seo et al. [14] and Wang et al. [15] proposed different solutions in ρ and λ domains, respectively, by referring to distortions of previous frames to improve one-pass RC, while Shen et al. [16] utilized two-pass RC to improve quality consistency. The second category uses a sliding window control method. Xu et al. [7]-[8] devised a window-level RC scheme for smooth quality and buffer control. The third category adopts the adaptive QP clipping control. Hu et al. [9] noticed the influence of QP clipping on quality consistency and adaptively set its range by considering frame complexity variation and buffer status. The last category resorts to prediction of QP increment. Sanz-Rodriguez et al. [10] predicted frame QP variation via radial basis network to control quality consistency in variable bit rate oriented video coding.

### B. Correlation Between Intra and Inter Coding

Existing works usually optimize bit allocation and QPs for inter frames or blocks [17], but neglect intra frames for its remarkable impact on overall performances [19]. To clearly manifest its importance, we can identify two issues. First, bit resources should be allocated in a trade-off manner between intra and inter frames to obtain overall optimization. Second, the distortion of intra frame can be propagated to inter frames because of prediction coding mechanism [20]-[21]. Hence, the correlation between intra and inter frame coding should be better analyzed [34] for improvements. Instead of pursuing

This work was supported by Shenzhen Science and Technology Plan Basic Research Project (JCYJ20190808161805519), Guangdong Basic and Applied Basic Research Foundation (2019A1515012031), Natural Science Foundation of China (61801303 and 62031013), Ministry of Science and Technology of China - Science and Technology Innovations 2030 (2019AAA0103501), and Open Projects Program of National Laboratory of Pattern Recognition (NLPR) (202000045). (*Corresponding author: Ge Li.*)

W. Gao, R. Wang, and G. Li are with the School of Electronic and Computer Engineering, Peking University, and also with Peng Cheng Laboratory, China (Email: gaowei262@pku.edu.cn, rgwang@pkusz.edu.cn, geli@pku.edu.cn).

S. Ma is with Institute of Digital Media, Peking University, Beijing, China (Email: swma@pku.edu.cn).

Q. Jiang is with School of Information Science and Engineering, Ningbo University, Ningbo, China (Email: jiangqiuping@nbu.edu.cn).

S. Kwong is with the Department of Computer Science, City University of Hong Kong (email: cssamk@cityu.edu.hk).

optimal solution for every frame, we will improve intra and inter frame coding trade-off to ameliorate quality consistency. In fact, determining intra frame QP (IQP) within each intra period is critical for intra and inter balanced coding. Then, establishing the relationship between quality consistency and IQP can lead to an improved solution.

A few IQP methods have been reported. Four different QPs were recommended in JVT-O079 [22] empirically for intra frame coding based on target bit rates. Yang et al. [23] also obtained IQP via bit rates, with parameters modeled based on edge magnitudes discriminately for different resolutions. Wang et al. [24] calculated IQP by utilizing entropy information of pixel-level gray scales and complexity measure of Intra16 DC mode, where different parameters were used for different resolutions. Hu et al. [4] refined IQP for H.264/SVC by evaluating available bandwidth, block-level variance of first intra frame and sum of absolute difference among frames. The HEVC reference software [25] adopted a modulated R-$\lambda$ model [26] to set IQP, where sum of absolute transformed difference (SATD) [19] was used as complexity measure. However, target bits for intra frames were empirically allocated, thus inaccuracy was introduced. In summary, existing IQP works are mostly either implicitly or explicitly for R-D gains [4], [22]-[27], rather than quality consistency. Obviously, they also actually lack clear definition on optimization goal and have too many empirical parameters. In addition, some IQP methods rely on the coding information in particular standards [4], [24], which cannot be used in other and future video coding standards. In [27], a learning-based method for adaptive quantization initialization was proposed, where optimal coding parameters are still selected with the criterion of R-D performance.

### C. Contributions and Organization

Unlike the traditional methods, we will give a learning-based method in rate control for quality consistency oriented video coding (LIRC-QC), and the main contributions are as follows:

(1) This is the first work to explore **a new optimization perspective for smooth quality**, which improves the coding trade-off between intra and inter frames. Contrarily, existing works have never discussed this problem explicitly and systematically, while it is critical for video coding, especially for better frame-level rate control.

(2) As far as we know, we are the first to **derive the quality consistency enhanced IQP**. Traditional methods neglect IQP influence, and usually set target frame distortion from previous information. This paper is also the pioneering effort to develop a learning-based method for consistency-oriented IQP.

(3) Moreover, we also devise efficient methods for ground truth refinement, equivalent feature acquisition, and adaptive QP clipping using real-time bandwidth and previous QPs.

**The first contribution gives a new perspective for quality consistency optimization, while the second and last give the general strategy and solutions to relevant detail problems, respectively.** The remainder is organized as follows. Section II studies the trade-off problem between intra and inter frame coding, and then gives a novel solution to to improve quality consistency via better intra frame QP prediction. Section III discusses the details of proposed method. Experimental results and comparisons are elaborated in Section IV, and finally the conclusion is drawn in Section V.

## II. CORRELATION ANALYSIS BETWEEN INTRA AND INTER FRAME CODING FOR QUALITY CONSISTENCY OPTIMIZATION

### A. Intra and Inter Frame Coding Trade-off

Fig. 1 illustrates the existed trade-off problem between intra and inter frame coding, which has not been fully explored. In Fig. 1 (a), the predictive mechanism in hybrid coding structure can propagate the distortion of referenced frames to the current frame. In Fig. 1 (b) to (e), five different IQPs, i.e., {14, 22, 30, 38, 46}, are tested for *Cactus* sequence, which are denoted by red circles, black stars, blue squares, cyan triangles and magenta diamonds, respectively, and frame QPs, coding bits, peak signal noise ratio (PSNR) and structural similarity metric (SSIM) frame quality are recorded. Coding results for the first intra period are collected, where R-$\lambda$ model based rate control algorithm is used in experiments and target bit rate is set as the consumed bits of fixed QP encoding with QP=27. Given bit rate constraint, intra frame with a larger QP will consume less bits and have lower quality, and subsequently inter frames can have more bits, smaller QPs and higher quality. However, due to distortion propagation, low quality intra frame will deteriorate inter coding. With this dependent relationship, high quality intra frame can similarly benefit the following inter coding. Hence, we can easily identify the balancing problem between intra and inter coding, and the important role of intra coding can be clearly seen from generated coding results.

Undoubtedly, different configurations can also draw the same conclusion due to the similar bit constraint condition and predictive coding dependency. Retaining the other parts of rate control, we can study whether there is an optimal IQP for the remarkable improvement of quality consistency.

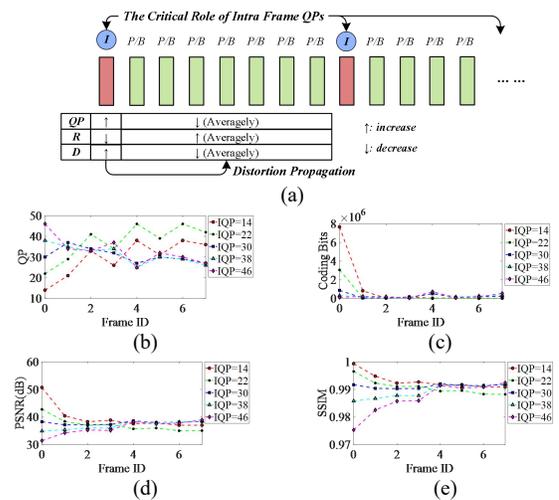

Fig. 1. Illustration of the existed coding trade-off problem between intra and inter frames. (a) Due to the constrained bit rate resources and the distortion propagation from predictive coding, the QP allocations for different frames become complicated for better rate control optimization. From (b) to (e), the frame-level variations of QP, coding bits, PSNR (dB) and SSIM values can be observed for the *Cactus* sequence, resulting in the motivation of determining the best intra frame QP for intra and inter frame coding trade-off.

There are mainly two metrics to evaluate quality fluctuation, i.e., the average absolute quality differences between adjacent frames, and the standard variance of frame quality. For convenience, StdvarPSNR and StdvarSSIM are defined as the standard variances of frame quality measured by PSNR and SSIM, respectively. Fig. 2 shows quality fluctuations using different IQPs, where encoding attempts are constrained by target bit rates from fixed frame QPs of {18, 24, 30, 36, 42} for *Traffic*, *Cactus*, *BasketballDrill*, *BlowingBubbles* and *KristenAndSara*. They are denoted by red circles, black stars, blue squares, cyan triangles and magenta diamonds, respectively. Without loss of generality, we have tested the low delay B coding structure with intra period size of 8, and GOP size of 4 on HEVC reference software HM-16.14. The other settings can also have the same conclusion that different IQPs will generate different results on quality consistency.

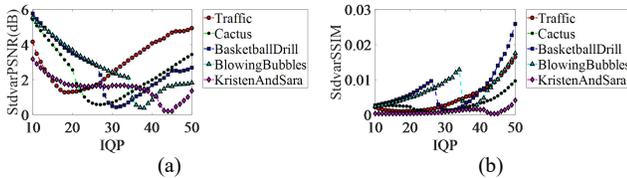

Fig. 2. Illustration of different degrees of frame quality fluctuations when using different intra frame QPs. The standard variances of PSNR and SSIM frame quality are denoted as StdvarPSNR (dB) and StdvarSSIM, respectively.

We can see that balanced intra and inter frame coding can effectively reduce quality fluctuations. Moreover, the intrinsic influential factor is actually the prediction-based redundancy removal process. Hence, different settings will have the same problem, and all the IQPs in different intra periods should be determined precisely. Additionally, the control of inter frame coding is implemented under the guidance of predicted intra frame QP and the adopted rate control algorithm framework.

### B. Formulation of Quality Consistency Improvement

Traditional quality consistency methods usually set target distortion for the current frame based on distortions of previous frames to control quality variations, and R-D-Q relationship should be modeled for bit and QP allocations. Because it is difficult to obtain R-D-Q model perfectly, they usually fail to obtain expected QPs. Moreover, the relationship is not clear between local frame smoothing and global quality consistency. In fact, the performance will also be significantly influenced by fluctuations of real-time available bit rates and video content. Hence, from this point of view, traditional methods can not easily get desirable performances.

Since selecting proper intra frame QPs is helpful to control fluctuation, the formulation can be given regarding quality consistency improvement by intra frame QP optimization. Under a particular rate control algorithm (*RCA*) with video content related factors (*VCF*), the score of QC (*SQC*) is:

$$SQC = F(IQP, RCA, VCF),  \quad (1)$$

where the function $F$ defines the relationship between quality consistency and all influential factors. The impact of IQP can be observed in above experiments in Section II-A, and the influences of RCA and VCF on quality consistency are also obvious. Consequently, IQP is proposed to be refined for the benefit of quality consistency. From the revealed correlation between intra coding and quality consistency, *SQC* can be ameliorated with desirable IQPs. Although there are many other critical issues to make encoders obtain smoother quality, these possible factors are difficult to model or observe. The optimal solution to QP allocation for all frames also requires a complex combinational optimization with consideration of bit consumption, distortion propagation and buffer control, etc. To simplify this complex problem, we do not pursue the theoretical optimal solution of quality consistency, but the improvement method using enhanced IQP. However, it is still not easy to model this interaction mechanism, and consequently we cannot directly derive the best solution for quality consistency.

Besides the weaknesses from lacking clear optimization goal and using empirical parameters, existing IQP methods still have not evaluated the effectiveness of influential factors. Hence, we need to firstly define quality consistency goal before executing IQP selection. As we have identified key factors that alters the score of quality consistency in Eq. (1), the achievement of the highest score could be formulated as:

$$OIQP = \arg\max_{IQP} SQC = \arg\max_{IQP} F(IQP, RCA, VCF), \quad (2)$$

where optimal IQP (OIQP) is obtained. It can be seen that Eq. (2) is very easily derived from Eq. (1), where IQP is defined as the major concern in optimization. However, due to the complex coding tools and video content, we can not deduce an effective calculative model to reflect this relationship for OIQP determination. In fact, this originates from the intractable trade-off problem between intra and inter frame coding. In virtue of sufficient data and model tuning, machine learning algorithms can work well on this kind of problems. Hence, we choose to give a learning-based approach:

$$OIQP = \min_{IQP} L(FS \mid RCA), \quad (3)$$

where $L$ is the loss function during learning to be defined and implemented later. Eq. (3) is given to illustrate the relationship between optimal intra frame QP and extracted useful feature set (FS) to help prediction performance, where other parts of rate control algorithm remain as fixed setting. Afterwards, quality consistency problem becomes OIQP prediction problem, where loss function will reach to its minimization. The overall process is regulated by rate control and extracted feature set. Eq. (3) also plays a critical role in the following learning-based formulation to predict the best coding parameters. Therefore, it can be seen that the intrinsic rationality for the descriptions of problem formulation is very clear, and the above three equations can also adequately illustrate the definition and transformation of the discussed problem.

The advantages of proposed learning-based approach can be highlighted from the following aspects. (1) This method can reduce the number of empirical parameters, which are usually required in traditional calculative models. (2) An effective solution can be derived for quality consistency optimization, which makes the complex problem become more manageable. By simply predicting coding parameters, the clear expected consistency goal becomes easily achievable.

## III. PROPOSED INTRA FRAME QUANTIZATION PARAMETER DETERMINATION AND RATE CONTROL METHOD

### A. Proposed Framework

Fig. 3 illustrates the framework of the proposed consistent quality oriented rate control method, namely LIRC-QC, which mainly includes three key steps. Due to the lack of machine learning databases for various prediction problems in video coding optimization, we prepare ground truths for optimal intra frame QPs in advance.

In the first step, by using different target bit rates to restrict encoding bandwidth under different intra frame QP attempts, the best QP leading to the smoothest quality is marked as ground truth for each sequence and each target bit rate, where selection criteria plays a critical role to guarantee the finally achieved video coding performance. In the second step, the high efficiency learning model for video coding actually has its own unique requirement on accuracy and complexity for practical encoder implementation, and therefore the elegant learning algorithm is demanded to be capable of achieving better video coding performance with allowable increase of computation complexity, rather than only prediction accuracy. Moreover, features should be extracted in an efficient manner. In the third step, trained model is applied to practical encoding process, and intra and inter frame coding can be optimized with the predicted intra frame QP, where both coding trade-off and consistent quality can be obtained under the fixed encoding configuration. More details of proposed method are discussed in the following paragraphs.

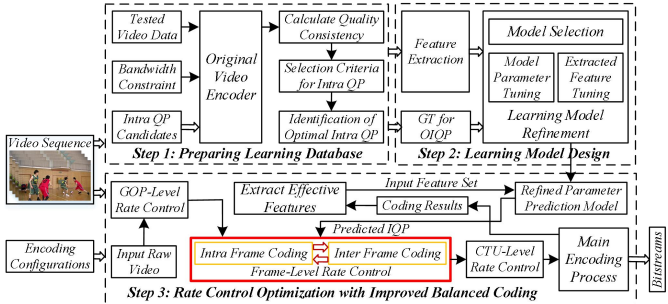

Fig. 3. Overall framework of proposed consistent quality oriented rate control.

### B. Identifying Optimal IQP Ground Truths

For learning data preparation, identifying ground truths for OIQP is the most important part. We should firstly clarify the evaluation criterion for quality consistency, and then sufficient experiments are required using different IQPs and bit rates. For fairness, we retain the other parts of rate control algorithm and only configure different IQPs.

Due to the characteristics of human visual perception, the visual comfort should be evaluated for a time duration. Instead of the average absolute difference between adjacent quality, SQC is defined as the reciprocal of frame-level quality standard variance. Hence, SQC and {StdvarPSNR, StdvarSSIM} are the evaluation pairs for quality consistency, which can be used equivalently. SQC can help visualize the comparison and the proposed removal method of unqualified candidates. Fig. 4 shows that different IQPs lead to different SQCs, and smooth shifting can also be seen between every adjacent IQPs. Here, the sequence meanings of different color and shapes are the same with those in Fig. 2.

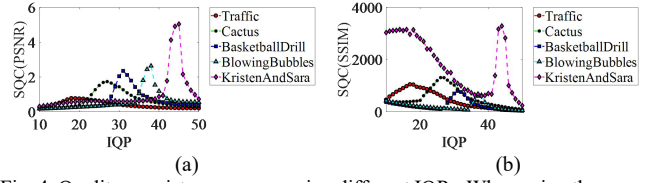

Fig. 4. Quality consistency scores using different IQPs. When using the score of quality consistency to evaluate the smoothness of video coding quality. (a) shows the PSNR-Based SQC fluctuations with different intra frame coding QPs, while (b) shows the SSIM-Based SQC fluctuations similarly.

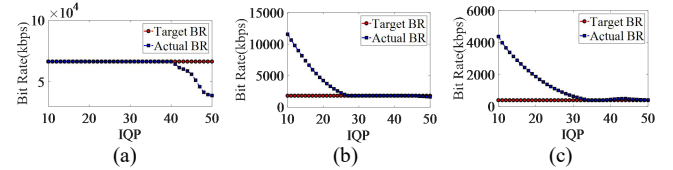

Fig. 5. Bit rate deviation phenomenon generated by different IQPs. (a) *Traffic*, (b) *BasketballDrill*, (c) *BlowingBubbles*.

Fig. 5 plots the achieved bit deviations by different IQPs. When IQP is smaller or larger, bit mismatch becomes much larger. One of these IQPs may have better SQC, but with lower bit rate accuracy. Hence, we propose to discard unsatisfactory IQPs with large bit deviation for the sake of overall rate control performance. Bit rate accuracy can be denoted as *BRA* to evaluate the absolute difference between target bit rates (*TBR*) and actual bit rates (*ABR*). A multiplicative penalty term (*PT*) is introduced to deselect unqualified IQPs with low BRAs:

$$PT = \begin{cases} 0, & \text{if } BRA > BRA\_threshold \\ 1, & \text{otherwise} \end{cases}, \quad (4)$$

where *BRA_threshold* is configured according to the practical requirements. Then, in a multiplication manner, SQC values are modulated as:

$$SQC \leftarrow SQC \times (1 - PT), \quad (5)$$

and refined IQPs can meet the specified BRA requirement. As we know, BRA is the basic goal of rate control optimization, where the achieved bit rate should be close to target bit rate. Thus, BRA value is always expected to be very high, otherwise rate control algorithm is not preferred.

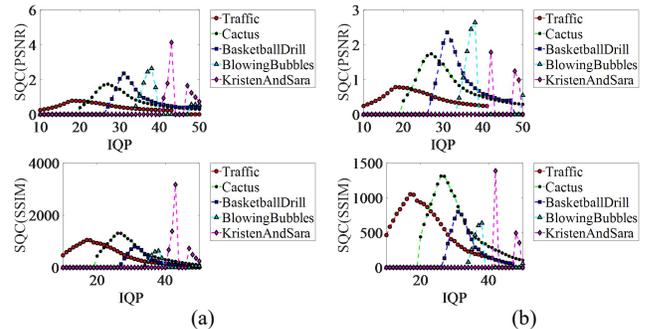

Fig. 6. Quality consistency scores with different IQP screening thresholds on bit rate accuracy based penalty term. (a) SQC (PSNR) and SQC (SSIM) with BRAC_Case1; (b) SQC (PSNR) and SQC (SSIM) with BRAC_Case2.

For simplicity, we denote the cases without BRA constraint as BRAC_Case0, the cases with BRA constraint of above 90% as BRAC_Case1, and the cases with BRA constraint of above 95% as BRAC_Case2. In this paper, we use the BRAC_Case2 threshold for testing. Fig. 6 shows results after screening. By eliminating unqualified IQPs, we can effectively deactivate the defective SQCs to guarantee the quality of OIQP candidates.

Since SQCs for different sequences have a large dynamic range, which is not convenient for comparison. Thus, we devise a normalization term using FixedQP results, where frames are encoded without QP offsets leading to low quality fluctuation [17]. Since FixedQP encodes every frame with a fixed value of QP, we can obtain smooth frame quality and a benchmark for rate control optimization, and multiple encoding attempts are needed to get the closest bit rate consumption to target bit rate. If all frames have the same quality, SQC will be $+\infty$. For each sequence, FixedQP results can be used as baselines. Hence, SQC ($i$) for the $i$-th IQP can be normalized as:

$$SQC(i) \leftarrow \frac{SQC(i)}{SQC(FixedQP)}, \quad (6)$$

where SQC (FixedQP) is from FixedQP using the same target bit rate constraint. In Fig. 6, fixed QPs, including {15, 18, 21, 24, 27, 30, 33, 36, 39, 42, 45 and 48}, are used for the first intra period and then we have 12 different target bit rates for each sequence. After normalization, SQCs become comparable for different sequences. OIQPs with the highest SQC (PSNR) and SQC (SSIM) are marked with red stars. IQP with the largest SQC is the best selection for quality consistency gains.

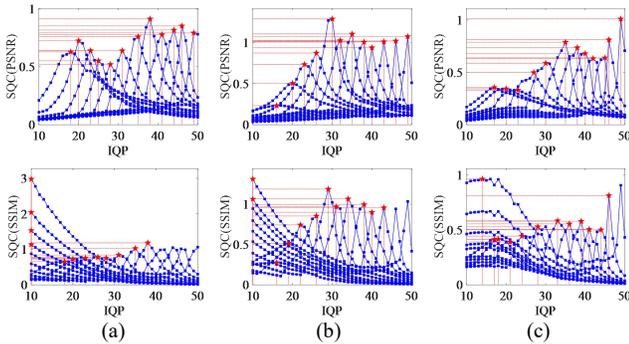

Fig. 7. OIQP identification without any bit rate accuracy-based screening (BRAC_Case0). (a) *BasketballDrill*, (b) *BlowingBubbles*, (c) *KristenAndSara*.

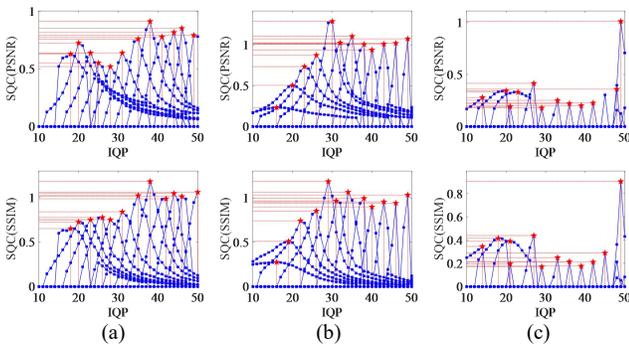

Fig. 8. OIQP identification with the proposed bit rate accuracy-based screening (BRAC_Case2). (a) *BasketballDrill*, (b) *BlowingBubbles*, (c) *KristenAndSara*.

There exist distinct unqualified IQPs in Fig. 7 (c) for PSNR based ground truths and Fig. 7 (a), (b) and (c) for SSIM based ground truths, where some IQPs with relatively lower bit rate accuracy lead to higher SQC results. By using the proposed BRAC_Case2 removal method, the corresponding subplots of Fig. 8 give acceptable OIQPs with high bit rate accuracy. Finally, compared with Fig. 7, the original bit mismatch cases with excessive pursuit of quality consistency can be effectively avoided. After establishing the IQP database, we can collect sufficient data samples, and then learning model and features can be optimized for the expected goal.

### C. Collecting Data Samples

We have tested all the recommended 18 video sequences. Without loss of generality, the settings with low delay B, intra period size of 8 and GOP size of 4 are tested. We have clarified that the proposed method is independent of the detailed coding settings. Moreover, to prepare learning data samples, we have executed simulations on high performance computing clusters for a few months to get the following ground truths and coding results. Therefore, due to limited pages, we only discuss the above exemplary coding setting to illustrate the nature of this method. For practical applications, videos usually have abrupt motions and scene changes, or video transmission channel is not perfect for signal retaining, short intra period size is suitable for these cases to guarantee video coding performance.

For each sequence, twelve different target bit rates are obtained by encoding only the first intra period using FixedQP. Rate control is the R-λ model-based method as in HM-16.14 [25]-[26]. For fair evaluations, the only difference lies in intra frame QPs, and different values from 10 to 50 are applied to achieve target bit rates. Hence, many encoding attempts are demanded for each OIQP ground truth.

Collected data of {*PeopleOnStreet*, *BasketballDrive*, *Cactus*, *ParkScene*, *BQMall*, *RaceHorsesC*, *BlowingBubbles*, *RaceHorses*, *Johnny*} are used as training set with 108 samples, while collected data of {*Traffic*, *BQTerrace*, *Kimono*, *BasketballDrill*, *PartyScene*, *BasketballPass*, *BQSquare*, *FourPeople*, *KristenAndSara*} are used as testing set with another 108 samples. The selection guideline is to divide each Class video sequence uniformly into training and testing sets. We did not have any prior knowledge for the learning configurations. Due to the huge computation complexity in data preparation, the number of samples cannot be too large, which also conforms with the light-weight preference of learning algorithm. Moreover, it should be noted that the overall learning data preparation, training and testing have been properly conducted to guarantee the generalization and efficiency of the learning model.

### D. Bridging OIQP and Prediction Model

For OIQP learning, besides high prediction accuracy, the most critical criterion is low complexity. Hence, support vector machine (SVM) [28] has been used in video coding as classifiers for fast mode decision [29]. Deep learning models may not be suitable for the above discussed problem, and the reason lies in the extremely high demand on data sample amount for deep neural network training and testing. Each ground truth takes a long time to prepare, and there are also no publicly databases for frame QP learning problem. Evidently, different from mode decision, it will definitely require much more time in preparing frame-level ground truths, where each sequence should be encoded too many times.

More specifically, we try 41 different IQPs to obtain a data sample for each sequence at each target bit rate. Hence, instead of using complex learning models, it is also wise to use a simple but practically effective model, where much less data samples are needed to reduce time consumption. Unlike the classifiers in mode decision, an effective regressor is required for OIQP prediction, which is ε-type SVM regression (ε-SVR) [28] with radial basis function (RBF) kernel. Training process finds the best hyperplane to acquire the model fitting parameters. Then, a linear model for samples can be constructed. We can define $<\cdot,\cdot>$ to get the dot product of two vectors, $x$ to represent input feature vector, $\varphi(\cdot)$ to represent RBF kernel, $\omega$ to represent weights to regress support vectors to the target hyperplane, and bias value is denoted as $b$. Finally, output $y$ is actually predicted IQP (PIQP). As shown in Eq. (3), optimal IQP is obtained by extracting effective features conditioned by a specific RCA, and this loss function can be equivalently reformulated as the following loss function in training and minimized:

$$\min_{IQP}\{L(FS|RCA)\} \Leftrightarrow \min_{IQP}\left\{J = \frac{1}{2}<\omega,\omega> + C\sum_{k=1}^{n}(\delta_k + \delta_k^*)\right\}, \quad (7)$$

where $k$ ($k=1, 2, ..., n$) is the data sample index, $\delta_k$ and $\delta_k^*$ are the slack variables which are both positive real values, and $C$ is trade-off parameter. Optimization will not penalize errors within ε-tube. Parameters are refined during training, including support vectors $\{s_t, t=1, 2, ..., SVN\}$ and the corresponding coefficients $\{\theta_t, t=1, 2, ..., SVN\}$, where $SVN$ is the number of support vectors. Eq. (7) clarifies the correlations between loss function definition in Eq. (3) and SVM model training process. The former finds optimal IQP to minimize quality fluctuation, while the latter optimizes model parameters by the training process of regressor. Finally, output PIQP is calculated as:

$$PIQP = y = \sum_{t=1}^{SVN} k_t \cdot e^{-\gamma \|s_t - x\|^2} + b, \quad (8)$$

where $\gamma$ is the variance-related gamma parameter in kernel. Training and testing are both simulated using LIBSVM [28], after which support vectors, corresponding coefficients and model parameters, i.e., $\gamma$ and $b$, can be achieved. Based on this model, PIQP can be used as IQP for smoother quality.

### E. Proposed Feature Analysis and Acquisition Method

The average and standard variance of QPs within the period are denoted as AvgQP and SvarQP, respectively. Except TBPP, the other seven features can be arbitrarily combined, and finally the best solution is determined as {TBPP, AvgQP, SvarQP} for both PSNR-Based and SSIM-Based optimization with RMSE of 2.986 and 3.154, respectively. The proposed method can effectively shrink IQP selection range from 0 to 51 to a very small range. Moreover, feature dimension and support vector number are also very small, which restrain model complexity. Compared with other texture-related features, this method extracting feature from bandwidth and QP information has the distinct advantage of lower storage and computation burden.

Although the effectiveness of AvgQP and SvarQP features has been validated, they are unavailable until the current intra period finishes. For simplicity and effectiveness, we propose an equivalent acquisition method for these two features, where the previous QP related information can be used as the alternative approximations. The proposed method is based on the fact that both average QP level and intra QPs are close for adjacent intra periods. For target bit rates generated by fixed QPs with increment of 3, both OIQP values and quality consistency scores are very close. Hence, close intra QPs will lead to close inter frame QP variations. Therefore, in most cases, every two adjacent intra periods will have similar evaluation values for QP variation. We also can evaluate the prediction accuracy of these features to verify rationality and guarantee the correctness of this alternative approach.

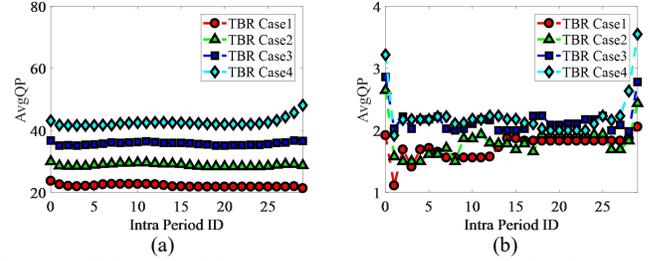

Fig. 9. AvgQP and SvarQP variations during real encoding for *ParkScene* video sequence.

TABLE I
PREDICTION ACCURACY OF SELECTED FEATURES

| Class | PSNR-Based Optimization | | | | SSIM-Based Optimization | | | |
|---|---|---|---|---|---|---|---|---|
| | AErr_AvgQP | AErr_SvarQP | AErr_AvgQP | AErr_SvarQP | AErr_AvgQP | AErr_SvarQP | AErr_AvgQP | AErr_SvarQP |
| A | 0.258 | 0.151 | 99.21 | 91.42 | 0.297 | 0.150 | 99.10 | 90.95 |
| B | 0.207 | 0.109 | 99.36 | 93.94 | 0.223 | 0.120 | 99.31 | 93.06 |
| C | 0.374 | 0.155 | 98.68 | 91.55 | 0.382 | 0.154 | 98.68 | 91.37 |
| D | 0.338 | 0.138 | 98.94 | 92.53 | 0.347 | 0.143 | 98.91 | 91.98 |
| E | 0.329 | 0.117 | 98.98 | 94.10 | 0.338 | 0.123 | 98.95 | 93.33 |
| Average | 0.301 | 0.134 | 99.03 | 92.71 | 0.317 | 0.138 | 98.99 | 92.14 |

Fig. 9 shows AvgQP and SvarQP variations during real encoding for ParkScene sequence, where smooth changes can be seen for these features. Hence, it is reliable to use previous QP information for current encoding. To further quantitatively evaluate accuracy, the following metrics are defined:

$$AErr = \frac{1}{IPN-1}\sum_{i=1}^{IPN-1}|AvgQP_i - AvgQP_{i+1}|, \quad (9)$$

$$RAcc = \frac{1}{IPN-1}\sum_{i=1}^{IPN-1}\left(1-\left|\frac{AvgQP_i - AvgQP_{i+1}}{AvgQP_{i+1}}\right|\right)\times 100\%, \quad (10)$$

where the total number of intra periods is $IPN$, while $AErr$ and $RAcc$ (%) give the absolute error and relative accuracy for AvgQP prediction, respectively. These can also be calculated for SvarQP similarly. By testing each sequence with four target bit rates, average $AErr$ and $RAcc$ results for AvgQP and SvarQP can be obtained. Table I shows that the accuracy for {AvgQP, SvarQP} can reach {99.03%, 92.71%} and {98.99%, 92.14%}, respectively, and absolute errors are also very low. Hence, the proposed feature prediction method can provide accurate equivalents, which ensures that the proposed model and feature extraction method can work efficiently.

### F. Bandwidth-Based QP Modulation

QP clipping can be commonly seen in video encoders [9], [25]. However, due to dynamic encoding contexts, it is actually difficult to give an optimal clipping strategy for robust gains. Traditional clipping methods for frames or blocks utilize frame complexity and buffer status information. Since the mismatch between OIQP and PIQP can lead to worse quality consistency and bit rate accuracy, we give a new clipping method based on real-time bandwidth and previous QPs.

More specifically, available bandwidth, i.e., target bit rates, for the current remaining encoding intra period is denoted as $ABW_{Cur}$, while $ABW_{All}$ represents available bandwidth for all intra periods in the total encoding sequence. IQPs in previous two periods are denoted as $IQP_{Prev\_1}$ and $IQP_{Prev\_2}$, respectively. Clipping methods usually give lower and upper boundaries, i.e., $minQP$ and $maxQP$. If $ABW_{Cur}$ is larger than $ABW_{All}$, we decrease IQP to be smaller than $mean(IQP_{Prev\_1}, IQP_{Prev\_2})$ for faster bandwidth consumption:

$$\begin{cases} \min QP = mean(IQP_{Prev\_1}, IQP_{Prev\_2}) - dQP \\ \max QP = mean(IQP_{Prev\_1}, IQP_{Prev\_2}) \end{cases}, \quad (11)$$

and otherwise,

$$\begin{cases} \min QP = mean(IQP_{Prev\_1}, IQP_{Prev\_2}) \\ \max QP = mean(IQP_{Prev\_1}, IQP_{Prev\_2}) + dQP \end{cases}, \quad (12)$$

where $dQP$ is set as 3 to ensure smooth changes of frame QPs [9]. Previous IQPs and real-time bandwidth are used for IQP adjustment. The first intra period has no clipping, while the second intra period only uses the first IQP. Following this particular rule, we ensure that IQPs modulated by previous QPs and bandwidth can efficiently avoid the excessive pursuit of quality consistency at the expense of bit rate inaccuracy.

### G. Overall Algorithm Flow

In **Algorithm 1**, we provide the general algorithm procedure of the proposed method to improve quality consistency, where the learning models and clipping methods are effectively used.

---

**Algorithm 1 Smooth Quality Oriented Intra Frame QP Determination**

This algorithm can provide the intra frame QP to obtain the frame-level smooth quality in video coding, where the target bit rate is given.

**Input:** Target bit rates for the current and all encoding intra periods: $ABW_{Cur}$ and $ABW_{All}$, Learning model with feature $\{TBPP\}$ predicting the first $IQP$: $LM_0$, Learning model with features $\{TBPP, AvgQP, SvarQP\}$ predicting other $IQPs$: $LM_1$, constant $dQP$=3.

**Output:** The best intra frame QP: $IQP_{best}$.

1: **if** IntraPeriodID=0
2:     $IQP=LM_0(ABW_{All})$, $IQP_{Prev\_1}=IQP$
3: **else if** IntraPeriodID=1
4:     $IQP=LM_1(ABW_{Cur}, AvgQP, SvarQP)$
5:     **if** $ABW_{Cur}>ABW_{All}$
6:       Clip $IQP$ within ($minQP=IQP_{Prev\_1}-dQP$, $minQP=IQP_{Prev\_1}$)
7:     **else**
8:       Clip $IQP$ within ($minQP=IQP_{Prev\_1}$, $minQP=IQP_{Prev\_1}+dQP$)
9:     **end if**
10: **else if** IntraPeriodID>1
11:     $IQP=LM_1(ABW_{Cur}, AvgQP, SvarQP)$
12:     $mIQP_{Prev}$=mean $(IQP_{Prev\_1}, IQP_{Prev\_2})$
13:     **if** $ABW_{Cur}>ABW_{All}$
14:       Clip $IQP$ within ($minQP=mIQP_{Prev}-dQP$, $minQP=mIQP_{Prev}$)
15:     **else**
16:       Clip $IQP$ within ($minQP=mIQP_{Prev}$, $minQP=mIQP_{Prev}+dQP$)
17:     **end if**
18: **end if**
19: Get $AvgQP$ and $SvarQP$ after encoding
20: Set IQP as the $IQP_{best}$

---

## IV. EXPERIMENTAL RESULTS

### A. Experiment Setup

To verify the performance of proposed quality consistency method, besides HM-16.14 [25], we implement the other two state-of-the-art quality consistency oriented RC methods, i.e., TIP16-Wang [15] and TIP13-Seo [14], and a state-of-the-art R-D performance oriented RC method, i.e., TBC19-Gao [27]. FixedQP uses sufficient encoding attempts to achieve the target bit rates. The encoding configurations are listed as Table II, while rate control is disabled for FixedQP simulations. Each video sequence is tested using four different target bit rates, which are obtained by applying FixedQP encoding to the first intra period with the same frame QPs {22, 27, 32, 37}.

TABLE II
EXPERIMENTAL CONFIGURATIONS

| Configurations | Values | Configurations | Values |
|---|---|---|---|
| Profile | Main | KeepHierarchicalBit | 2 |
| Coding Structure | Low Delay B | LCULevelRateControl | 1 |
| MaxCUWidth | 64 | RCLCUSeparateModel | 1 |
| MaxCUHeight | 64 | MaxDeltaQP | 0 |
| MaxPartitionDepth | 4 | MaxCuDQPDepth | 0 |
| IntraPeriod | 8 | RDOQ | 1 |
| GOPSize | 4 | RDOQTS | 1 |
| FastSearch | 1 | RCForceIntraQP | 0 |
| SearchRange | 64 | IntraQPOffset | -1 |
| RateControl | 1 | InitialQP | 0 |

### B. Quality Consistency

We compare the consistency results using StdvarPSNR (dB) and StdvarSSIM for different methods. In Table III, compared with HM-16.14, the proposed method can averagely reduce quality fluctuation by 37.2% (from 1.748 dB to 1.098 dB) on StdvarPSNR and by 45.1% (from 0.0051 to 0.0028) on StdvarSSIM. For simplicity, it is noted that TBC19-Gao results are directly from the given results in [27]. We can see that our method can improve quality consistency significantly.

TABLE III
QUALITY FLUCTUATIONS FOR DIFFERENT RATE CONTROL METHODS

| Class | HM16.14 | | TIP16-Wang | | TIP13-Seo | | TBC19-Gao | | Proposed LIRC-QC | | FixedQP | |
|---|---|---|---|---|---|---|---|---|---|---|---|---|
| | Stdvar PSNR | Stdvar SSIM | Stdvar PSNR | Stdvar SSIM | Stdvar PSNR | Stdvar SSIM | Stdvar PSNR | Stdvar SSIM | Stdvar PSNR | Stdvar SSIM | Stdvar PSNR | Stdvar SSIM |
| A | 2.146 | 0.0062 | 3.587 | 0.0179 | 1.809 | 0.0038 | 1.331 | 0.0042 | 0.789 | 0.0016 | 0.156 | 0.0006 |
| B | 1.422 | 0.0037 | 1.853 | 0.0080 | 1.624 | 0.0040 | 0.976 | 0.0036 | 0.817 | 0.0025 | 0.385 | 0.0017 |
| C | 2.046 | 0.0069 | 2.346 | 0.0121 | 2.763 | 0.0082 | 1.933 | 0.0083 | 1.590 | 0.0043 | 0.527 | 0.0019 |
| D | 2.151 | 0.0082 | 2.297 | 0.0113 | 2.691 | 0.0104 | 2.312 | 0.0143 | 1.435 | 0.0050 | 0.526 | 0.0026 |
| E | 0.973 | 0.0007 | 1.524 | 0.0016 | 1.674 | 0.0009 | 1.354 | 0.0022 | 0.860 | 0.0006 | 0.229 | 0.0003 |
| **Average** | **1.748** | **0.0051** | **2.321** | **0.0102** | **2.112** | **0.0055** | **1.581** | **0.0065** | **1.098** | **0.0028** | **0.365** | **0.0014** |

### C. R-D Performance

Although quality consistency gains have been validated, R-D performance should also be maintained. For fairness, the four different target bit rates are tested for each sequence, and then Bjøntegaard delta BR (BD-BR), Bjøntegaard delta PSNR (BD-PSNR) [30], and Bjøntegaard delta SSIM (BD-SSIM) can be calculated from these four different R-D results.

Table IV presents R-D performances for different methods. BD-SSIM results are from the SSIM-Based OIQP optimization. Compared with HM-16.14, the proposed method has -8.128% BD-BR reductions, 0.298 dB BD-PSNR gains, and 0.0007 BD-SSIM gains, respectively. More gains can be achieved over the quality consistency oriented methods, i.e., TIP16-Wang and TIP13-Seo. Although R-D performance is not originally taken into account in learning model and ground truth establishment, the consistency-oriented IQP can also help coding efficiency. Moreover, these R-D gains also demonstrate that the proposed method can greatly ameliorate the trade-off between intra and inter frame coding to improve the overall coding performances.

TABLE IV
R-D PERFORMANCES OVER HM-16.14

| Class | TIP16-Wang | | | TIP13-Seo | | | TBC19-Gao | | | Proposed LIRC-QC | | | FixedQP | | |
|---|---|---|---|---|---|---|---|---|---|---|---|---|---|---|---|
| | BD-BR | BD-PSNR | BD-SSIM | BD-BR | BD-PSNR | BD-SSIM | BD-BR | BD-PSNR | BD-SSIM | BD-BR | BD-PSNR | BD-SSIM | BD-BR | BD-PSNR | BD-SSIM |
| A | 22.326 | -0.757 | -0.0051 | 3.085 | -0.151 | 0.0003 | -7.174 | 0.390 | 0.0006 | -12.307 | 0.471 | 0.0009 | -10.020 | 0.384 | 0.0001 |
| B | 17.922 | -0.290 | -0.0013 | 6.398 | -0.171 | -0.0002 | -19.723 | 0.672 | 0.0016 | -7.888 | 0.236 | 0.0001 | -8.052 | 0.259 | -0.0002 |
| C | 8.462 | -0.340 | -0.0017 | 15.381 | -0.601 | -0.0018 | -18.978 | 1.097 | 0.0073 | -6.207 | 0.258 | 0.0014 | -3.100 | 0.112 | 0.0013 |
| D | 7.012 | -0.289 | -0.0008 | 16.044 | -0.740 | -0.0025 | -15.554 | 0.898 | 0.0099 | -3.260 | 0.160 | 0.0012 | 0.154 | 0.016 | 0.0009 |
| E | 18.558 | -0.553 | -0.0004 | 25.319 | -0.403 | 0.0000 | -15.973 | 0.853 | 0.0007 | -10.980 | 0.367 | 0.0001 | -6.069 | 0.213 | -0.0001 |
| Average | 14.856 | -0.446 | -0.0018 | 13.245 | -0.413 | -0.0008 | -15.480 | 0.782 | 0.0040 | -8.128 | 0.298 | 0.0007 | -5.417 | 0.197 | 0.0004 |

### D. Bit Rate Achievement

As a basic goal of rate control, bit consumption should be accurately controlled to meet the target requirement. Bit rate accuracy can measure the mismatches between target and actual bit rates. The results are compared in Table V, which are the average results from four different bit rates. It can be seen that the proposed methods can have higher accuracy results than all the other quality consistency oriented methods, which are averagely 99.549% and 99.408%, respectively. This also validates the effectiveness of proposed bit rate accuracy-based removal method for establishing IQP ground truths.

TABLE V
COMPARISON ON BIT RATE ACCURACY (%)

| Class | HM-16.14 | TIP16-Wang | TIP13-Seo | TBC19-Gao | Proposed LIRC-QC (PSNR) | Proposed LIRC-QC (SSIM) | FixedQP |
|---|---|---|---|---|---|---|---|
| A | 98.389 | 96.462 | 96.491 | 99.525 | 99.455 | 99.017 | 97.327 |
| B | 98.492 | 99.327 | 95.552 | 99.957 | 99.859 | 99.794 | 94.030 |
| C | 97.814 | 98.258 | 98.018 | 99.919 | 99.888 | 99.857 | 96.982 |
| D | 98.772 | 97.070 | 92.684 | 99.942 | 99.845 | 99.841 | 97.737 |
| E | 97.640 | 98.532 | 56.638 | 99.863 | 98.697 | 98.533 | 95.928 |
| Average | 98.221 | 97.930 | 87.877 | 99.841 | 99.549 | 99.408 | 96.401 |

### E. Buffer Occupancy Control

Satisfactory control on buffer occupancy can be beneficial to decrease overflow and underflow occurrences when managing compressed bit streams, and thus can eliminate missing blocks and frames. When the missing pixels are compensated, more distortions will be generated. Hence, the stable buffer status can guarantee the achievement of high quality encoded videos. As we know, rate control algorithm can be optimized to achieve better buffer management.

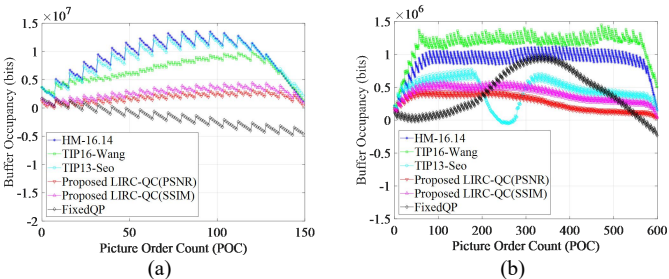

Fig. 10. Comparison on buffer occupancy. (a) *Traffic* (target bit rates: 12720.63 kbps), (b) *BQSquare* (target bit rates: 1956 kbps).

To avoid the limited scope of analyses, the existing buffer evaluation methods [7], [15], [17], [19] do not use fixed buffer sizes to count the buffer breakdowns. Instead, buffer status is recorded to deduce the possibility of abnormal situations. If buffer level lower than zero lasts longer, it will more easily encounter underflow. If buffer level higher than the others lasts longer, it will more easily encounter overflow.

Using this principle, Fig. 10 compares the buffer occupancy for different methods, where the proposed method has the least possibility to encounter buffer management problems. FixedQP method easily have underflows, while the other rate control methods tend to encounter overflows. Therefore, the proposed method can maintain better buffer occupancy status, and thus probably provide much robust high video quality.

### F. Scene Change

For scene change videos, short intra period size, e.g., Intra Period Size=8 as discussed in this paper, will become effective for better video coding performance. Thus, for full evaluation of quality consistency oriented rate control algorithms, it is necessary to test scene change videos. Although the proposed method is not so explicitly designed for scene change videos, our machine learning based method can better comprehensively consider the content change and coding dependency among all intra and inter frames.

As is well known, abrupt changes in videos greatly influence the behavior of encoder and its performances. We test the synthesized videos in [21] with frequent changes. Table VI and Fig. 11 demonstrate that the proposed method can obtain smaller fluctuations than the other rate control methods, where these StdvarPSNR and StdvarSSIM results are obtained from the PSNR-Based and SSIM-Based optimization approaches, respectively. From Table VII, Table VIII and Fig. 12, more results can show the achieved remarkable gains by the proposed method on R-D performance, bit rate accuracy and buffer control over the existing quality consistency oriented methods.

TABLE VI
QUALITY FLUCTUATIONS FOR SCENE CHANGE VIDEOS

| Sequence | HM-16.14 | | TIP16-Wang | | TIP13-Seo | | TBC19-Gao | | Proposed LIRC-QC | | FixedQP | |
|---|---|---|---|---|---|---|---|---|---|---|---|---|
| | Stdvar PSNR | Stdvar SSIM | Stdvar PSNR | Stdvar SSIM | Stdvar PSNR | Stdvar SSIM | Stdvar PSNR | Stdvar SSIM | Stdvar PSNR | Stdvar SSIM | Stdvar PSNR | Stdvar SSIM |
| MixedB1 | 2.403 | 0.0111 | 2.959 | 0.0164 | 2.412 | 0.0105 | 2.007 | 0.0098 | 1.725 | 0.0077 | 1.362 | 0.0069 |
| MixedB2 | 1.633 | 0.0046 | 2.390 | 0.0163 | 2.093 | 0.0042 | 0.972 | 0.0025 | 0.898 | 0.0014 | 0.670 | 0.0009 |
| MixedC | 2.734 | 0.0195 | 3.053 | 0.0264 | 2.971 | 0.0184 | 2.386 | 0.0094 | 2.305 | 0.0070 | 1.311 | 0.0027 |
| MixedD | 2.563 | 0.0156 | 2.938 | 0.0152 | 2.946 | 0.0132 | 2.375 | 0.0157 | 1.870 | 0.0049 | 1.294 | 0.0016 |
| Average | 2.333 | 0.0127 | 2.835 | 0.0186 | 2.605 | 0.0116 | 1.935 | 0.0093 | 1.699 | 0.0053 | 1.159 | 0.0030 |

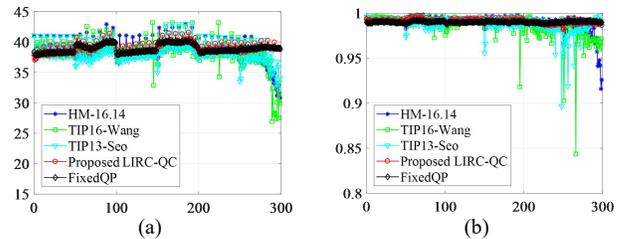

Fig. 11. Comparison on quality consistency for scene change videos. X-axis: picture order count, Y-axis: PSNR/SSIM. (a) MixedB2 (target bit rates: 5204 kbps, PSNR-based), (b) MixedD (target bit rates: 1454.7 kbps, SSIM-based).

### G. More Results and Analyses

We denote the proposed method without QP clipping as "LIRC-QC-NC", and Table IX shows the compared results from different aspects between LIRC-QC-NC and the proposed method. It can be seen that these two schemes can have similar quality consistency on both StdvarPSNR and StdvarSSIM. However, the proposed adaptive QP clipping method can also generate noticeable gains on the R-D performance and bit rate controlling accuracy.

TABLE VII
R-D PERFORMANCES FOR SCENE CHANGE VIDEOS OVER HM-16.14

| Sequence | TIP16-Wang | | | TIP13-Seo | | | TBC19-Gao | | | Proposed LIRC-QC | | | FixedQP | | |
|---|---|---|---|---|---|---|---|---|---|---|---|---|---|---|---|
| | BD-BR | BD-PSNR | BD-SSIM | BD-BR | BD-PSNR | BD-SSIM | BD-BR | BD-PSNR | BD-SSIM | BD-BR | BD-PSNR | BD-SSIM | BD-BR | BD-PSNR | BD-SSIM |
| MixedB1 | 10.879 | -0.324 | -0.0020 | 9.360 | -0.263 | -0.0003 | -16.413 | 0.701 | 0.0014 | -12.092 | 0.369 | 0.0014 | -12.859 | 0.396 | 0.0010 |
| MixedB2 | 11.538 | -0.306 | -0.0021 | 14.225 | -0.356 | -0.0002 | -17.571 | 0.742 | 0.0026 | -15.925 | 0.410 | 0.0012 | -18.573 | 0.497 | 0.0014 |
| MixedC | 31.083 | -0.851 | -0.0037 | 8.835 | -0.389 | 0.0001 | -13.123 | 0.852 | 0.0069 | -9.505 | 0.433 | 0.0055 | -3.918 | 0.168 | 0.0054 |
| MixedD | 13.658 | -0.496 | -0.0002 | 15.191 | -0.700 | 0.0000 | -19.182 | 1.130 | 0.0131 | -7.024 | 0.348 | 0.0042 | -4.578 | 0.241 | 0.0042 |
| Average | 16.789 | -0.494 | -0.0020 | 11.903 | -0.427 | -0.0001 | -16.572 | 0.856 | 0.0060 | -11.136 | 0.390 | 0.0031 | -9.982 | 0.326 | 0.0030 |

TABLE VIII
BIT RATE ACCURACY FOR SCENE CHANGE VIDEOS (%)

| Sequence | HM-16.14 | TIP16-Wang | TIP13-Seo | TBC19-Gao | Proposed LIRC-QC (PSNR) | Proposed LIRC-QC (SSIM) | FixedQP |
|---|---|---|---|---|---|---|---|
| MixedB1 | 97.207 | 98.321 | 89.220 | 99.712 | 99.587 | 99.359 | 95.265 |
| MixedB2 | 98.917 | 99.310 | 98.896 | 99.925 | 99.946 | 99.935 | 93.042 |
| MixedC | 97.694 | 97.816 | 95.592 | 99.994 | 99.850 | 99.851 | 94.961 |
| MixedD | 98.645 | 96.370 | 97.662 | 99.950 | 99.968 | 99.935 | 94.825 |
| Average | 98.169 | 97.093 | 96.627 | 99.895 | 99.909 | 99.893 | 94.893 |

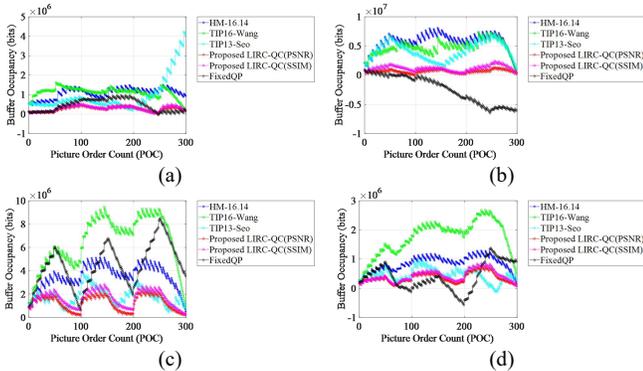

Fig. 12. Buffer occupancy for scene change videos: (a) *MixedB1* (target bit rates: 976.944 kbps); (b) *MixedB2* (target bit rates: 13224.65 kbps); (c) *MixedC* (target bit rates: 11699.4 kbps); (d) *MixedD* (target bit rates: 3778.85 kbps).

Fig. 13 shows the visual quality comparison. It can be seen that the proposed method can effectively produce high quality reconstructed videos with less compression artifacts under the same target bit rate constraint, indicating better rate control and video coding performances.

Most of existing quality assessment metrics cannot perform universally well for different video sources, degradation types and distortion levels, etc. Therefore, by integrating different kinds of quality metrics with support vector regression, the video multimethod assessment fusion (VMAF) metric [31]-[32] has raised attention for better perceptual experience evaluation. We use VMAF to replace PSNR and SSIM, and denote StdvarVMAF as the standard variance of frame-level VMAF quality, and then calculate the BD-VMAF result.

Table X gives the comparisons on BD-VMAF and StdvarVMAF between the proposed LIRC-QC method and the HEVC reference software. We can find that the proposed PSNR and SSIM based method can reduce quality fluctuations by 47.1% (from 6.462 to 3.416) and 47.8% (from 6.462 to 3.376), respectively. Moreover, the BD-VMAF can have 2.851 and 2.757 gains, respectively. Additionally, Fig. 14 gives the frame-level VMAF fluctuation results. Hence, all these results demonstrate that the proposed method can also perform much better when using VMAF for quality evaluation. Subjective demos can be seen in the link provided as [35].

TABLE IX
PERFORMANCE EVALUATION FOR QP CLIPPING

| Class | Proposed LIRC-QC-NC | | | | | | Proposed LIRC-QC | | | | | |
|---|---|---|---|---|---|---|---|---|---|---|---|---|
| | BRA (PSNR) | BRA (SSIM) | BD-BR | BD-PSNR | BD-SSIM | Stdvar PSNR | Stdvar SSIM | BRA (PSNR) | BRA (SSIM) | BD-BR | BD-PSNR | BD-SSIM | Stdvar PSNR | Stdvar SSIM |
| A | 99.177 | 98.877 | -11.353 | 0.437 | 0.0008 | 0.780 | 0.0016 | 99.455 | 99.017 | -12.307 | 0.471 | 0.0009 | 0.789 | 0.0016 |
| B | 99.825 | 99.750 | -7.613 | 0.237 | 0.0001 | 0.789 | 0.0025 | 99.859 | 99.794 | -7.888 | 0.236 | 0.0001 | 0.817 | 0.0025 |
| C | 99.810 | 92.536 | -5.439 | 0.224 | -0.0020 | 1.503 | 0.0044 | 99.888 | 99.857 | -6.207 | 0.258 | 0.0014 | 1.590 | 0.0043 |
| D | 99.833 | 99.772 | -3.060 | 0.148 | 0.0012 | 1.395 | 0.0052 | 99.845 | 99.841 | -3.260 | 0.160 | 0.0012 | 1.435 | 0.0050 |
| E | 98.528 | 98.280 | -8.969 | 0.302 | 0.0000 | 0.767 | 0.0006 | 98.697 | 98.533 | -10.980 | 0.367 | 0.0001 | 0.860 | 0.0006 |
| Average | 99.434 | 97.843 | -7.287 | 0.270 | 0.0000 | 1.047 | 0.0029 | 99.549 | 99.408 | -8.128 | 0.298 | 0.0007 | 1.098 | 0.0028 |

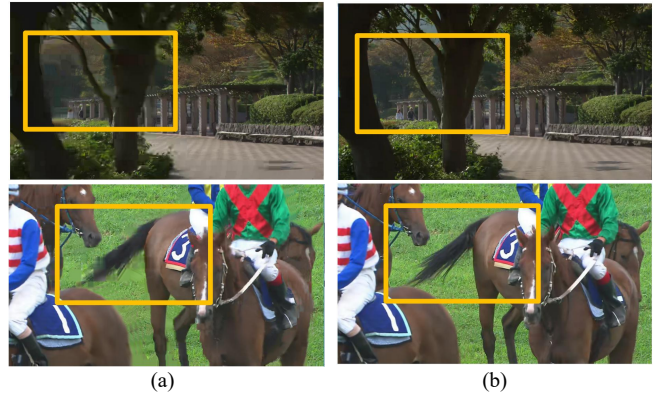

Fig. 13. Visual quality comparison for reconstructed videos. The two rows show results for *ParkScene* (target bit rates: 966.696 kbps) and *RaceHorsesC* (target bit rates: 3824.5 kbps). (a) HM-16.14, (b) Proposed LIRC-QC (above row: PSNR-Based optimization, below row: SSIM-Based optimization).

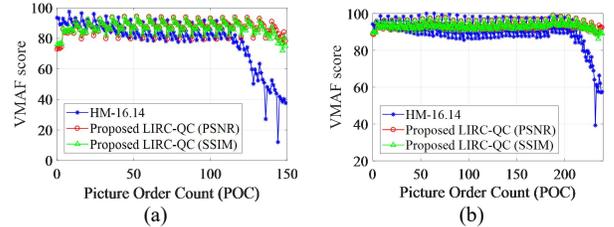

Fig. 14. Comparisons on VMAF based quality consistency. (a) *Traffic* (target bit rates: 5272.29 kbps), (b) *ParkScene* (target bit rates: 6716.856 kbps).

TABLE X
VMAF-BASED PERFORMANCE COMPARISON

| Class | HM-16.14 | Proposed LIRC-QC (PSNR) | | Proposed LIRC-QC (SSIM) | |
|---|---|---|---|---|---|
| | Stdvar VMAF | BD-VMAF | Stdvar VMAF | BD-VMAF | Stdvar VMAF |
| A | 9.520 | 4.434 | 2.884 | 4.263 | 2.959 |
| B | 6.118 | 2.397 | 3.404 | 2.410 | 3.289 |
| C | 6.653 | 2.985 | 3.990 | 2.873 | 4.043 |
| D | 7.362 | 3.103 | 4.245 | 3.043 | 4.325 |
| E | 2.658 | 1.336 | 2.556 | 1.198 | 2.265 |
| Average | 6.462 | 2.851 | 3.416 | 2.757 | 3.376 |

All these experimental results can effectively validate that the proposed method can perform much better than the other state-of-the-art rate control methods. Moreover, it is also strongly believed that the same approach can be easily extended to other encoding configurations, e.g., different intra period sizes and predictive coding structures, and it is expected to have significant performance improvements effectively.

V. CONCLUSION

To effectively improve quality consistency of compressed videos, this paper proposes to balance intra and inter frame coding. First, we give a new perspective and framework to obtain consistent quality in video coding, where the critical importance of intra frame coding is highlighted. Second, we propose the learning-based intra frame QP determination method, including effective solutions to establish the learning ground truths, feature refinement and the equivalent feature acquisition method. Finally, an adaptive QP clipping method based on real-time bandwidth and previous coding information is also proposed, which is beneficial to bit rate controlling accuracy. Experimental results demonstrate the effectiveness of proposed method in achieving better coding performances. The main contribution of this work can be summarized as the proposed novel perspective and method to optimize quality consistency by ameliorating the coding trade-off between intra and inter frames. This work can be effectively applied to quality consistency oriented video encoder optimization. For the future work, the theoretical optimal solution to quality consistency optimization can be derived and advanced learning paradigms can be devised effectively for this problem.